# Fabrication and Characterization of $Fe_{100-x}Ni_x$ Nanoparticles in the Invar Region

F. H. Rawwagah[1], A-F. Lehlooh[1], S. H. Mahmood[2*], S. Mahmoud[1],
A-R. El-Ali[1], M. R. Said[1], I. Odeh[1], I. Abu-aljarayesh[1]

[1] Physics Department, Yarmouk University, Irbid, Jordan
[2] Physics Department, The University of Jordan, Amman, Jordan ([*]corresponding author)
E-mail: s.mahmood@ju.edu.jo

**Abstrac**

In this work, $Fe_{100-x}Ni_x$ nanoparticle systems in the invar region ($x$ = 29, 32, and 37) were prepared by the method of chemical co-precipitation. X-Ray Diffraction (XRD) patterns confirmed the coexistence of both bcc and fcc phases for $x$ = 29 and 32. However, only fcc phase was observed for $x$ = 37. Scanning Electron Microscope (SEM) images indicated that the particle size is ~ 120-500 nm, and is almost independent of $x$. Mössbauer spectroscopy (MS) on the prepared nanoparticles indicated the development of a paramagnetic phase characteristic of the low-spin (antitaenite) fcc phase, in addition to the magnetic components characteristic of the bcc phase and the high-spin fcc phases. Since it is not possible to resolve the low- spin fcc phase from the high-spin fcc phase in XRD patterns, MS proves to be an effective tool for studying the prepared nanoparticle systems.

**Key words**: Chemical co-precipitation, Mössbauer Spectroscopy, Superparamagnetic relaxations, γ´-(antitaenite) phase.

# Fabrication and Characterization of Fe$_{100-x}$Ni$_x$ Nanoparticles in the Invar Region

F. H. Rawwagah[1], A-F. Lehlooh[1], S. H. Mahmood[2*], S. Mahmoud[1],
A-R. El-Ali[1], M. R. Said[1], I. Odeh[1], I. Abu-aljarayesh[1]

[1] Physics Department, Yarmouk University, Irbid, Jordan
[2] Physics Department, The University of Jordan, Amman, Jordan ([*]corresponding author)
E-mail: s.mahmood@ju.edu.jo

1. Introduction

Iron nickel alloys in the invar region (65 – 70 at. % Fe) are of special interest to scientists and engineers due to their interesting anomalous thermal and magnetic properties [1-10]. Invar alloys demonstrated several anomalies such as low thermal expansion coefficient at room temperature, instability of ferromagnetism, low Curie temperature and saturation magnetizations, large high field susceptibilities, and relatively flat magnetization versus temperature. They also exhibited anomalies in their atomic volume, elastic modulus, and heat capacity. The Invar (derived probably from Invariant) effect was discovered by Guillaume in 1897 [1]. The observed effects result from the anomalously low thermal expansion of these alloys, which makes them of high dimensional stability under changing temperature. Therefore, they are widely used in precision mechanical systems in many different industries as well as opto-mechanical engineering applications. Iron nickel alloys near Invar (36 at. % Ni) concentration were investigated widely since the discovery of their low thermal expansion in 1897. The coefficient of thermal expansion (CTE) of the Invar 36 alloy at room temperature is approximately $1\times10^{-6}$ K$^{-1}$. Of course, like most mechanical properties CTE varies with temperature. However, Invar's CTE is the lowest of any metal. Actually, CTE for Invar is much below the CTEs of the most broadly used metals, aluminum and steel, and varies slowly with temperature which makes it suitable for some applications involving significant temperature changes. The low CTE of Invar alloys is very favorable for building different systems that require temperature invariant dimensions such as laser cavities. Longitudinal mode separation of a laser cavity depends very sensitively on the distance between the mirrors of the cavity. This distance must be maintained fixed with a tolerance of up to the wavelength of the laser (~ 0.5 μm for visible light lasers), which demands materials with near zero thermal expansion coefficient at room temperature.

In the Fe- rich regin, Fe-Ni alloys exist in the α-phase (bcc), whereas in the Ni-rich region they exist in the γ-phase (fcc), and a mixture of the two phases was observed for alloys with intermediate concentrations [2,4,5]. In addition to these two ferromagnetically ordered phases, Mössbauer spectroscopy demonstrated the presence of a paramagnetic phase which was attributed to superparamagnetic relaxations or to a low spin γ´-phase [4, 5, 10-12]. However, the composition of the alloy at which this component appears, and its relative abundance in the alloy were not identical in the literature, demonstrating the sensitivity of this component to the preparation method. The magnetic anomalies in Fe-Ni alloys prepared by chemical co-precipitation [4] and by arc melting [5] near the Invar 36 concentration were thus previously investigated by Mössbauer spectroscopy. In this concentration range, magnetic and structural phase transitions were observed, and the low spin component appeared. This component was attributed to superparamagnetic clusters [5], and partly to the low spin γ´-phase (antitaenite) [7].

In this work we have prepared Fe-Ni nanoparticles using the chemical co-precipitation method, and investigated their structural and magnetic properties using X-ray diffraction, Electron Microscopy, and Mössbauer spectroscopy.

2. Experimental

The alloys $Fe_{71}Ni_{29}$, $Fe_{68}Ni_{32}$, and $Fe_{63}Ni_{37}$ were prepared using chemical co-precipitation. Proper amounts of Iron-oxide $Fe_2O_3$ powder for each sample was dissolved in a proper amount of Nitric acid ($HNO_3$) with the aid of heat. A proper amount of Nickel-acetate $Ni(CH_3CO_2)_2$ salt was dissolved in distilled water. The solution for each sample was then added at once to a clean beaker together to form a homogeneous mixture of the two solutions. A pH of ~ 7 for the new solution was reached by adding ammonia solution. The metal ions in the mixture were precipitated by adding a proper amount of Sodium oxide $Na_2O$ solution. The precipitate was washed by distilled water several times until pH reading of ~7 was reached. The sample was then dried in air for ~12 hours at 100 ºC. The dry metal carbonate mixture was powdered and placed in a catalytic fixed bed flow reactor through which air was passed at a flow rate of 150 $cm^3$/min. and a temperature of 300°C for one hour. The produced homogeneous metal oxide mixture was then reduced in a hydrogen gas atmosphere at 300°C for 3 hours, and the resulting alloy powder was cooled to room temperature by an electric fan for about an hour, under helium atmosphere.

Scanning Electron Microscopy (SEM) was used to investigate the morphology of the particles and the particle size distributions for the prepared samples. X-ray diffraction for the powder samples is performed using a standard θ-2θ diffractometer with Co-K$_α$ (λ = 1.79025 Å) radiation. Mössbauer Spectroscopy in transmission geometry was conducted using a standard constant acceleration Mössbauer spectrometer, and the spectra were collected over 512 channels. Isomer shifts were measured relative to the centroid of the α-Fe spectrum at room temperature, and the spectra were fitted using a software based on least squares analysis.

3. Results and Discussion

Scanning electron microscope (SEM) images for the samples $Fe_{100-x}Ni_x$ ($x$ = 29, 32, 37) show that the samples consist of uniform distributions of almost spherical particles with diameters ranging from ~ 120 – 500 nm (Figs. 1). The particle size distribution does not appear to depend on the Ni content in this concentration range.
XRD patterns (Fig. 1) indicate that the sample with $x$ = 29 consists of two phases; the α-bcc FeNi phase and the γ-fcc phase. The γ-phase increases with increasing Ni concentration until $x$ = 37 where the sample becomes a pure fcc phase. The lattice parameter of the bcc phase is (2.85 ± 0.01) Å, while that of the fcc phase is (3.55 ± 0.01) Å. These are only slightly lower and higher than the lattice parameters for bcc Fe (2.87 Å) and fcc Ni (3.52 Å), respectively. The small difference in lattice parameters could be associated with the fact that Ni atom has a smaller size than Fe atom. In the concentration range of this study, the difference in lattice parameters for the three samples is within experimental uncertainty.

The crystallite size is measured from the broadening of the diffraction lines using the Scherrer equation:

$$D = \frac{K\lambda}{\beta \cos\theta}$$

Here $K$ is a constant (0.94), $\lambda$ is the wave length of X-ray (1.79 Å), and $\beta$ is the width of the diffraction peak at half maximum. The measured crystallite size for the samples is (35 ± 5) nm. The crystallinity of the samples does not appear to depend appreciably on the Ni concentration, which indicates that the intrance of Ni into the bcc Fe, or that of Fe into fcc Ni, does not influence the crystal structure appreciably.

Mössbauer spectra for the different samples are shown in Fig. 3 and the hyperfine parameters obtained from fitting the spectra with the appropriate components are listed in table 1. The spectrum for the sample with $x = 29$ was fitted with two magnetic sextet components and a central singlet. The component with $B_{hf} = 34.3$ T is associated with the bcc FeNi phase, and that with $B_{hf} = 29.2$ T is associated with the fcc FeNi phase. These values are consistent with those reported earlier for these phases in alloys with $x = 30$ prepared by arc melting [5] and by chemical co-precipitation [7]. The central singlet with width of 0.82 mm/s and relative intensity of 15% is also consistent with that observed for the sample with $x = 30$ prepared by arc melting [5]. However, this singlet is appreciably broader than that with width of 0.40 mm/s observed in the sample prepared by co-precipitation. This broadening could be associated with superparamagntic relaxations in small particles. Such a relaxation in superparamagnetic clusters which results in broadening of the singlet was reported earlier [5, 13].

The spectrum for the sample with $x = 32$ was also fitted with two magnetic sextets corresponding to the bcc and fcc phases, and a broad central singlet that could be associated with superparamagnetic relaxations. The relative intensity corresponding to the fcc phase grows at the expense of the bcc phase (table 1), which is consistent with the behavior of the relative intensities of the two phases in the XRD patterns (FIG. 2). However, the relative intensity of the singlet drops down to about half its value for the previous sample, which could be associated with a reduction of the proportion of the small superparamagnetic particles in this sample.

The spectrum of the sample with $x = 37$ was fitted with two sextets with hyperfine fields of 30.0 T and 16.3 T, in addition to a relatively narrow central singlet with width of 0.49 mm/s. The two low field sextets were previously reported for the alloy with $x = 35$ prepared by arc melting [5], and were assigned to fcc phase. This assignment is consistent with the pure phase observed in the XRD pattern for this sample. The peaks of the magnetic component with $B_{hf} = 16.3$ T coalesce into a broad magnetic sextet. This component was shown in our previous work to be associated with superparamagnetic relaxations. Thus, this component in the present work is attributed to superparamagnetic relaxations in nanoparticles that haves grown somewhat larger than in the previous two samples. A similar component was detected in a previous work [6]. The narrow singlet is consistent with that reported earlier for FeNi powders prepared by chemical co-precipitation [7] and is assigned to the paramagnetic γ´-low spin (antitaenite) phase developing in regions with $x < 30$. The different heat treatment used in this study, however, could be responsible for the difference in the relative proportion of this component, and the concentration at which it occurred.

## 4. Conclusions

The method of chemical co-precipitation proved to be suitable for the preparation of invar FeNi nanoparticles with approximately uniform particle size distribution. The prepared nanoparticle systems demonstrated anomalous structural and magnetic phase transitions in a narrow range of elemental concentrations around the invar stoichiometry. The prepared FeNi nanoparticles by chemical co-precipitation demonstrated different magnetic properties and phase transitions than the bulk system prepared by arc melting [5]. The antitaenite phase developed in the concentration region around the invar stoichiometry (36 at.% Ni), and superparamagnetic relaxations associated with the nanoparticles were observed up to that concentration, while such relaxations were not observed in alloys with $x > 30$ prepared by arc melting as demonstrated by our previous work [5]. The variations of the magnetic phases and their relative proportions at different values of $x$ could indicate the sensitivity of these phases to the preparation procedure and the particle size distribution.

**Acknowledgement**

This work is supported by a generous grant from the Scientific Research Support Fund (SRSF), Jordan.

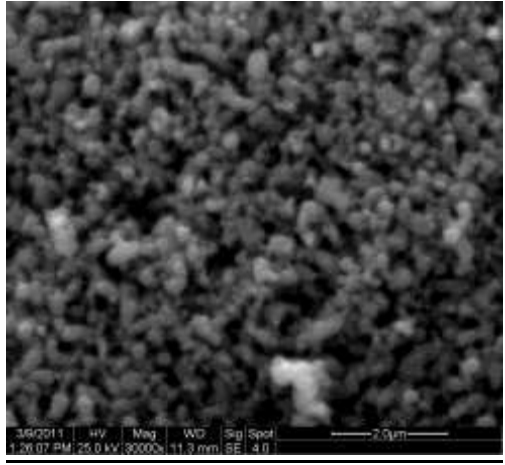

(a)

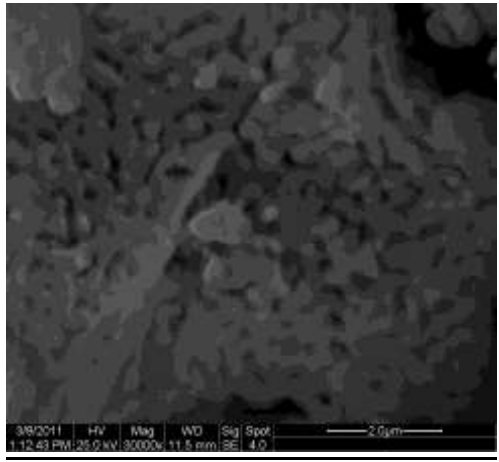

(b)

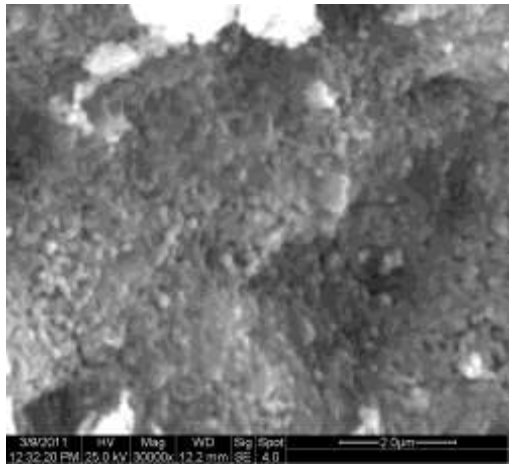

(c)

Fig. 1: SEM images for $Fe_{100-x}Ni_x$ with (a) $x = 29$, (b) $x = 32$, and (c) $x = 37$.

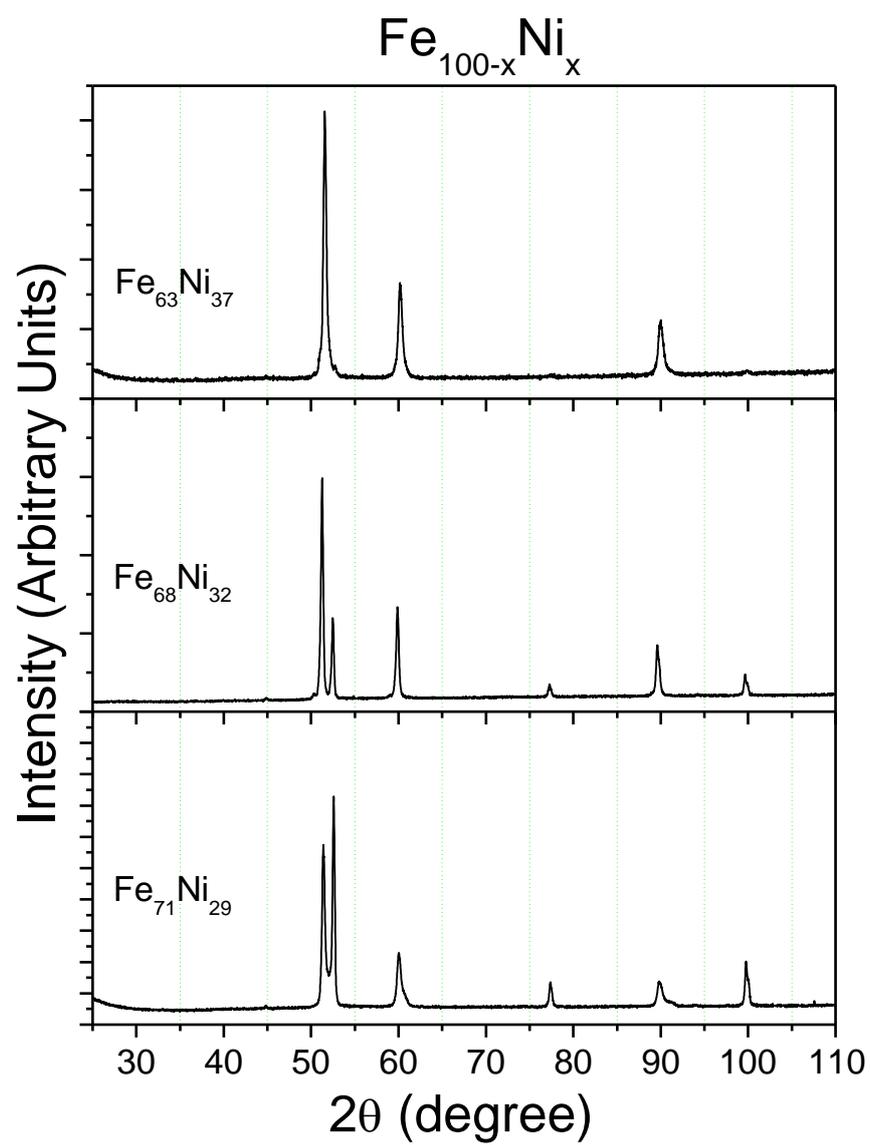

Fig. 2: XRD patterns of Fe$_{100-x}$Ni$_x$ nanoparticles.

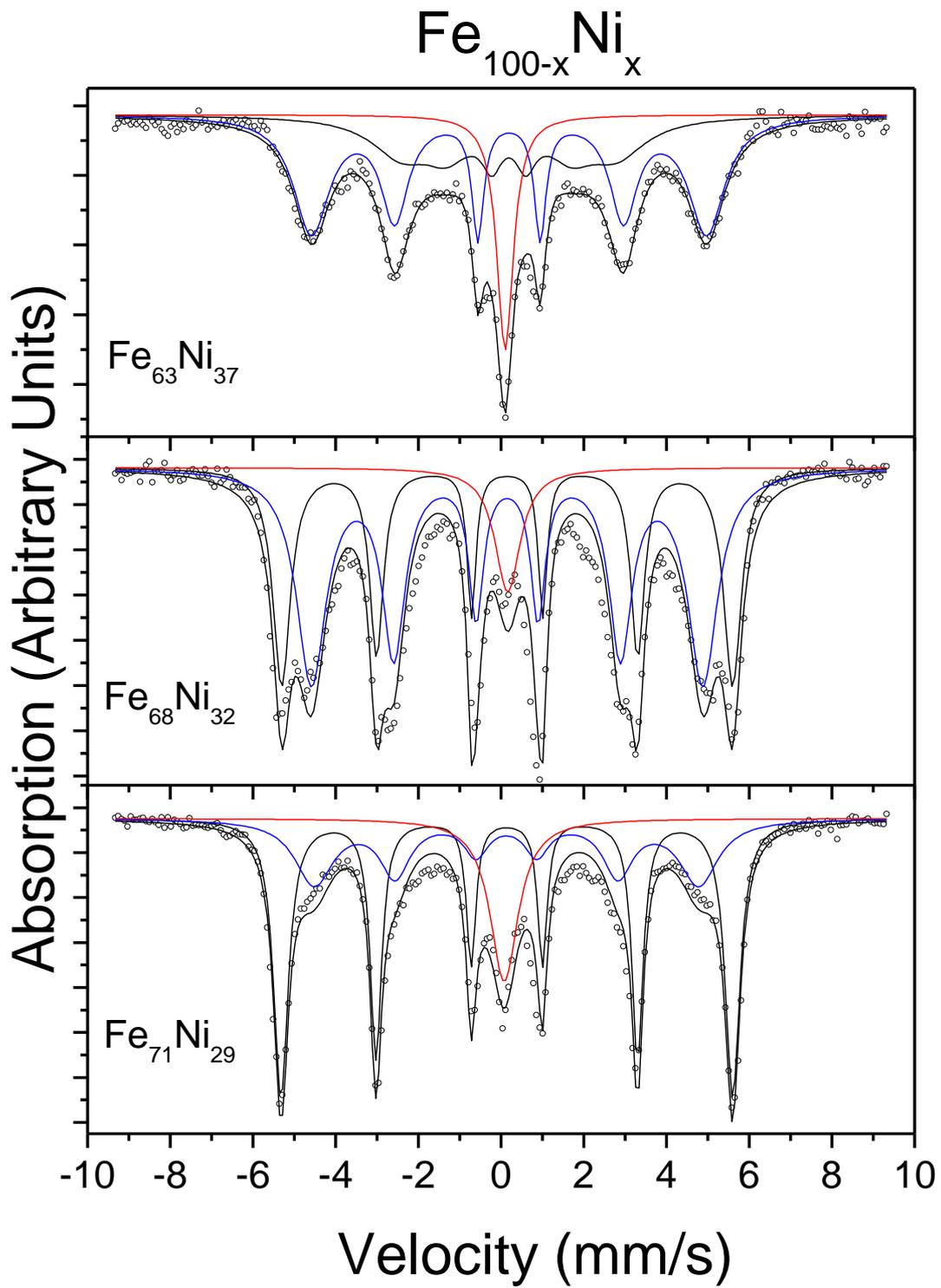

Fig. 3: Mössbauer spectra for the $Fe_{1-x}Ni_x$ system ($x$ = 0.29, 0.32, and 0.37).

Table 1: Isomer shift (IS) in mm/s, hyperfine magnetic field ($B_{hf}$) in Tesla (T), relative intensity (I), and width (W) of the inner lines in mm/s of Mössbauer spectra for the $Fe_{100-x}Ni_x$ system.

|  | $x = 29$ | $x = 32$ | $x = 37$ |
|---|---|---|---|
| $IS_1$ | 0.01 | 0.02 | 0.07 |
| $IS_2$ | 0.00 | 0.02 | 0.06 |
| $IS_3$ (Singlet) | -0.05 | 0.04 | -0.03 |
| $B_{1hf}$ | 34.3 | 34.2 | 30.0 |
| $B_{2hf}$ | 29.2 | 29.7 | 16.3 |
| $I_1$ | 0.53 | 0.35 | 0.57 |
| $I_2$ | 0.32 | 0.58 | 0.32 |
| $I_3$ (Singlet) | 0.15 | 0.07 | 0.13 |
| $W_1$ | 0.27 | 0.25 | 0.34 |
| $W_2$ | 0.71 | 0.43 | 0.58 |
| $W_3$ (Singlet) | 0.82 | 0.74 | 0.49 |